"Cigar" Fermi surface as a possible requisite for superconductivity in iron-based superconductors

S. V. Borisenko[1], A. N. Yaresko[2], D. V. Evtushinsky[1], V. B. Zabolotnyy[1], A. A. Kordyuk[1,3], J. Maletz[1], B. Büchner[1], Z. Shermadini[4], H. Luetkens[4], K. Sedlak[4], R. Khasanov[4], A. Amato[4], A. Krzton-Maziopa[5], K. Conder[5], E. Pomjakushina[5], H-H. Klauss[6], E. Rienks[7]

[1]*Leibniz-Institute for Solid State Research, IFW-Dresden, D-01171 Dresden, Germany*

[2]*Max-Planck-Institute for Solid State Research, D-70569 Stuttgart, Germany*

[3]*Institute for Metal Physics, 03142 Kyiv, Ukraine*

[4]*Laboratory for Muon Spin Spectroscopy, Paul Scherrer Institute, CH-5232 Villigen PSI, Switzerland*

[5]*Laboratory for Developments and Methods, Paul Scherrer Institute, CH-5232 Villigen PSI, Switzerland*

[6]*Institut für Festkörperphysik, TU Dresden, D-01069 Dresden, Germany*

[7]*Helmholtz-Zentrum Berlin, BESSY, D-12489 Berlin, Germany*

**Recently discovered *A*-Fe-Se (*A* - alkali metal) materials have questioned the most popular theories of iron-based superconductors because of their unusual electronic structure [1]. Controversial photoemission data taken in the superconducting state [2-7] are in conflict with highly magnetic state seen by neutron-, $\mu$SR-spectroscopies and transport/thermodynamic probes [8-10]. These results lead to suggestions to consider all iron-based materials as originating from Mott-insulators or semiconductors, thus once again raising the question of close relation between the cuprates and Fe-based superconductors [e.g. 2]. Here we study electronic and magnetic properties of $Rb_{0.77}Fe_{1.61}Se_2$ ($T_c$ = 32.6 K) in normal and superconducting states by means of photoemission and $\mu$SR spectroscopies as well as band structure calculations. We demonstrate that the puzzling behavior of these novel materials is the result of separation into metallic (~12%) and insulating (~ 88%) phases. Only the former becomes superconducting and has a usual electronic structure of electron-doped FeSe-slabs. Our results thus imply that the antiferromagnetic insulating phase is just a byproduct of Rb-intercalation and its magnetic properties have hardly any relation to the superconductivity. Instead, we find that also in this, already third class of iron-based compounds, the key ingredient for superconductivity is a certain proximity of a van Hove singularity to the Fermi level. These findings set the direction for effective search of new superconducting materials.**

Natural attempts to modify the properties of the simplest iron-based superconductor FeSe led to the synthesis of a new class of *A*-Fe-Se compounds [1]. Considerable enhancement of the critical temperature ($T_c$) up to 33 K stimulated active research in this field. It was found that *A*-Fe-Se are long range magnetically ordered systems with unprecedented for iron pnictides and chalcogenides high magnetic moment (3.3 $\mu_B$) and Néel temperature ($T_N$=535 K) [8-10], their resistivity is universally insulating down to 100 K and phase diagram implies that the parent state for all superconducting samples is an antiferromagnetic semiconductor [11]. This is in sharp contrast to the results of angle-resolved photoemission (ARPES) studies. All of them clearly demonstrate metallicity of the electronic structure, though without a consensus regarding the Fermi surface (FS) itself [2-7]. First, electron-like sheets were found in K-Fe-Se at the corners of the Brillouin zone (BZ) [4] with no evidence for electronic states at the Fermi level near the Γ-point. Alternatively, already three types of the electron-like Fermi surfaces were detected in Tl-K-Fe-Se, Tl-Rb-Fe-Se in Refs. 5-7. Finally, two FS sheets, in the center and in the corner of the BZ, were found in 100% superconducting samples of K-Fe-Se and Cs-Fe-Se [3] or in K-Fe-Se samples with rare inclusions of insulating regions [2]. The Luttinger count was found to correspond to the nominal doping ranging from 11% [4] to 31.5% [5] of an electron per Fe, thus implying a 100% metallic phase. On the contrary, it was concluded that the superconductor develops from a semiconductor as a result of nontrivial doping [2]. In all cases the angle-resolved photoemission data have been taken in the superconducting state, thus leaving the question of the normal state electronic structure, necessary to understand the superconductivity, open.

In Fig.1 we show the results of our ARPES study of the Rb-Fe-Se sample in the normal state ($T$ = 100 K). The experiments have been carried out using the synchrotron radiation from the BESSY (Helmholtz-Zentrum Berlin) storage ring. The FS maps in panels a) and b) are recorded using the light with mutually perpendicular vectors of linear polarization to detect the symmetry of the electronic states at the Fermi level. We observe only two types of the Fermi surfaces, in the center and in the corners of the BZ with typical for iron-based superconductors polarization dependence of the intensity. The structure along the momentum cuts through the high-symmetry points is in agreement with the band structure calculations, though the size of our corner centered FS is much smaller than in the calculations of the stoichiometric $RbFe_2Se_2$ (see Fig. 1f). The bandwidth renormalization is of the factor of 3, as can be estimated by comparing the binding energies of the prominent features in the spectra (Fig. 1 c, e) with those of the pile-ups of the dispersing curves from Fig. 1d. The typical Fermi velocity corresponding to the corner-centered FS is about 0.3 eVÅ. Thus, by ARPES we clearly see a metal, at least at $T$ = 100 K, and its electronic structure is described by LDA calculations of $RbFe_2Se_2$. We derive the charge carrier concentration from the Fermi surface area to be about 0.15 electrons per iron atom. This is obviously less than in stoichiometric $RbFe_2Se_2$ (0.5 e$^-$/Fe), but more than in $Rb_2Fe_4Se_5$ (0 e$^-$/Fe), the composition which is very close to the average composition of our single crystals ($Rb_{0.77}Fe_{1.61}Se_2$) determined by EDX. One can thus conclude that the metallic parts seen by ARPES cannot represent the whole sample, which is in contrast to the majority of earlier ARPES studies.

To elucidate the magnetic properties of the $Rb_{0.77}Fe_{1.61}Se_2$ crystal, zero field (ZF) and weak transverse field (wTF) muon spin rotation/relaxation (μSR) measurements have been performed at the Swiss Muon Source at the Paul Scherrer Institute, Switzerland. In such experiments, 100% spin polarized positive muons are implanted into the specimen and the time dependence of the spin polarization of the ensemble is obtained from their parity violating asymmetric decay. If the muon spin is subject to a perpendicular magnetic field $B$ (either external or internal) it precesses with a Lamor frequency $\omega_L$ = $\gamma_\mu B$, with $\gamma_\mu$ = 851.69 MHz/T being the gyromagnetic ratio of the muon. Since the muon is a local probe, the paramagnetic volume fraction of a sample can be determined by applying a weak transverse magnetic field and subsequently measuring the amplitude of the corresponding precession signal. In Fig. 2a) the result of such a measurement is shown. At high temperatures 100% of the muon spins precess in the externally applied magnetic field of $B_{ext}$ = 5 mT. Below a sharp transition at $T_N$ = 535 K large internal magnetic fields $B_{int}$ are present due to the magnetic ordering of the Fe moments. Correspondingly the precession amplitude of the $B_{ext}$ = 5 mT signal, i.e. the paramagnetic volume fraction, decreases. At low temperatures, approximately 88% of the sample is magnetically ordered, while ~12% of the crystal stays paramagnetic down to lowest temperatures. This phase separation into magnetic and non-magnetic volumes is consistently observed also in other $A_y$Fe$_{2-x}$Se$_2$ crystals by local probe techniques [12-16]. The low temperature magnetic structure is very well ordered as evidenced by the long-lived oscillation observed in the ZF µSR spectra shown in Fig. 2b). The corresponding internal field at 2 K of $B_{int}$(2K)=2.929(7) T at the muon site is also well defined in the direction of the crystallographic c-axis as µSR oscillations are observed when the muon spin is aligned perpendicular, but not when it is parallel to the c-axis. These observations are consistent with the long range ordered block spin antiferromagnetic state observed for the Fe vacancy ordered structure [17-19].

Our band structure calculations of Rb$_2$Fe$_4$Se$_5$ show that the ordered Fe vacancies lead to the magnetic insulator (Fig. 2c,d) with a magnetic moment of 3.5 µ$_B$ and the energy gap of 0.25 eV. The macroscopic phase separation into non-magnetic metallic and antiferromagnetic insulating parts is in agreement with the TEM study [20] and is supported by the fact that we do not observe any other signal at the Fermi level (replica), typical for magnetic metallic systems. In order to observe the metallic part we had to find a particular place at the sample surface, otherwise exhibiting significantly suppressed intensity at the Fermi level, typical for the insulators with such energy gap. On the other hand, these metallic areas have to be of considerable size as they are perfectly described by coherent electronic states of RbFe$_2$Se$_2$ (see Fig.1). This means that the iron vacancies, when disordered (Fig. 1g), tend to create clusters with the excess of positive charge which compensate the macroscopic electron-doped regions where the concentration of 0.15 electrons per Fe atom is achieved.

We now want to understand whether the superconductivity is related to the metallic or magnetic regions. To look for a possible coupling of the magnetic and superconducting order parameter, the ZF µSR spectrum at 40 K has been recorded to compare with the 2 K data in Fig.2b. Within the extremely small error bar of the measurement neither the magnetic volume fraction nor the size of the magnetic order parameter [$B_{int}$(40K) = 2.937(7) T] is changed above and below the superconducting transition at $T_c$ = 32.6 K. It should be noted that a clear reduction of the magnetic order parameter below $T_c$ has been observed in many other Fe-based superconductors exhibiting a microscopic coexistence and competition of the two forms of order [21-24]. On the other hand, a microscopic coexistence without an apparent competition has been reported for FeSe under hydrostatic pressures above P = 1.9 GPa [25]. Therefore the absence of a reduction of the magnetic order parameter below $T_c$ is not a proof, but may suggest that the 88% magnetic volume of the investigated Rb$_{0.77}$Fe$_{1.61}$Se$_2$ crystal is not superconducting in accordance with the insulating behavior expected for the Rb$_2$Fe$_4$Se$_5$ vacancy ordered structure (Fig. 2c) and features of the ARPES data. The remaining 12% paramagnetic volume clearly exhibit superconductivity as shown earlier for the very same single crystal [13]. Comparison of electronic structure below and above $T_c$ clearly demonstrates that the superconductor evolves from the metallic part of the sample (Figs. 1, 3). All features remain unchanged, except for the superconducting energy gap, which opens on both types of the electron-like Fermi surfaces (Fig. 1 a,b). In contrast to LiFeAs [26], we found no evidence of significant gap anisotropy. Typical bending-back of the dispersion is observed (Fig. 3d) and the typical size of the superconducting energy gap can be inferred directly from the peak of the energy-distribution curve (EDC) measured at $k_F$ (Fig. 3c).

With band dispersion and momentum dependence of the superconducting gap at hand, we can calculate electronic properties in the superconducting state and compare obtained values from

ARPES and µSR. A response to the external magnetic field is characterized by the London penetration depth, $\lambda_L$, which in the clean limit depends only on the distribution of the Fermi velocity and superconducting gap [27]. As follows from ARPES data, the Fermi velocity exhibits a weak anisotropy and the band parameters for the electron pockets at the BZ corners are $v_F$ = 0.27 eVÅ, $k_F$ = 0.25 Å$^{-1}$. Neglecting the small FS sheet in the center of the BZ, we obtain from ARPES data $\lambda_L$ = 250 nm at temperatures well below $T_c$, matching the value of 258 nm, extracted from µSR data [13]. Taking into account scattering effects, resulting in finite mean free path of electrons, one would get a larger value for $\lambda_L$ [28]. Temperature dependence $\lambda(T)$, measured by µSR, can be well fitted with isotropic superconducting gap of 7.7 meV [13], very close to the binding energy of the 8 meV peak in the EDC, discussed above (see Fig. 3c) and the gap value found by optical spectroscopy [29]. Such a good agreement between the experiments confirms that the observed ARPES spectra originate from the superconducting part of the sample.

Our results suggest that the studied material is macroscopically separated into magnetic insulating (88 %) and superconducting metallic phases (12 %). Now we can try to find out which ingredient of the electronic structure of the metallic parts is responsible for the superconductivity with the critical temperature of 32.6 K. The most peculiar feature of the electronic structure of RbFe$_2$Se$_2$ is the dispersions near the center of the BZ. In Fig. 4a,c we show the results of band structure calculations along the $\bar{X}$-$\bar{\Gamma}$-$\bar{X}$ of the 2D BZ (corresponding to NTN direction in Fig. 1f) by summing the spectral weight ascribed to dispersion curves for each $k_z$-point and using the same color scale as in Figs. 1,3 for ARPES data. The energy location of the two-dimensional $d_{xy}$-band in the zone center, seen as the narrowest parabolic hole-like feature with no $k_z$-dispersion in Fig. 4a,c, is not in agreement with ARPES experiments on A-Fe-Se and LiFeAs [26, 30], therefore we neglect its presence here. The hole-like bands appear to interact with electron-like band near $E_f$ resulting in a pronounced singularity. This singularity leads to an enhanced peak in the density of states provided the non-parabolic character of the band discussed in Ref. 30 is taken into account. Such interplay between hole- and electron-like bands near the Fermi level is observed in all known to us superconducting iron pnictides and chalcogenides. This behavior is schematically shown in the insets to Fig. 4. Two typical cases seem to be associated to the occurrence of a high $T_c$. Either an electron-like band approaches the Fermi level from an unoccupied side and crosses it for most $k_z$ - values or a hole-like band remains below the Fermi level and crosses it only within the narrow $k_z$ - interval. Remarkably, the optimally hole-doped 122 materials may host both cases (Fig. 4b,e) because of the rich "propeller" structure near the corner of the BZ [31]. In these two cases such a band structure results in small cigar-like Fermi surfaces (Fig. 4 d-f). The whole set of corresponding singularities, if considered separately for each $k_z$, is located below $E_F$ in the energy interval defined by the $k_z$ - dispersion (typically of the order of 10 meV). This substantially facilitates the matching of the energy of, e.g., the bosonic mode and thus the superconductivity itself.

That the band-edge type of van Hove singularity is universally present in Fe-pnictides has been noticed earlier [30]. The new aspects here are that the singularities are stronger because of the presence of an electron-like counterpart and that the band crosses $E_F$ with a finite $k_z$ -dispersion. At this, the most of the singularities should be located below the Fermi level, i.e. the "cigars" should be short if hole-like and long if electron-like. The latter condition seems to be essential, since the electron-overdoped Co-122, where the electron-like band also makes a small electron pocket at the center of the BZ, is not a superconductor. There the electron-like band has much larger $k_z$ -dispersion because of different energetic location of As $p$ and Se $p$ bands.

Our data taken in the normal state thus imply that there are no "parent" insulating or semiconducting materials of iron-based superconductors and the superconductivity arises from the metallic phase, namely from $RbFe_2Se_2$. Existence of stoichiometric non-magnetic materials like FeSe and LiFeAs together with the present case of Rb-Fe-Se clearly show that the key features are to be sought in fermiology of these compounds. The magnetic phase appears to be another competing order, hardly relevant for superconductivity. Obviously, if purely metallic $A_xFe_2Se_2$ with bulk doping level of 0.15% per iron atom was synthesized, another superconducting material with considerable critical temperature could be obtained. More generally, certain proximity of the van Hove singularity to the Fermi level is required. Possible realization of this resonant condition can be achieved in quasi 2D compounds with small in-plane FS (top of the band should be close to $E_F$) and low Fermi velocity in $k_z$ direction (tuning to boson frequency). These requirements can be formulated in terms of the "cigar" shaped Fermi surfaces, either short or long depending on their topology.

**Acknowledgements**

We acknowledge useful discussions with Igor Morozov and support of Rolf Follath at the beamline. This work was supported by the DFG priority program SPP1458, grants BO1912/3-1 and BO1912/2-2.

Figure 1. **Metal. a,b)** Fermi surface maps measured with horizontal and vertical polarizations of the incoming light with photon energy of 80 eV. Dashed lines are the FS contours with weaker intensity. **c,e)** Photoemission intensity along the cuts with $k_x=0$ Å$^{-1}$ and $k_x=1.1$ Å$^{-1}$ which are close to the PTP high symmetry direction in the BZ. **d)** Band structure of the stoichiometric RbFe$_2$Se$_2$ along the PTP direction for comparison with ARPES data. **f)** The PTN section of the calculated Fermi surface of RbFe$_2$Se$_2$, chosen as most representative for comparison with ARPES data. T-point is located half-way between Γ and Z points of the BZ. **g)** Schematic top-view of FeSe slabs with disordered iron vacancies. Large patches of FeSe without Fe vacancies are shown as shaded areas.

Figure 2. **Insulator. a)** Paramagnetic volume fraction as determined by weak transverse field (5 mT) μSR. Note the magnetic transition at $T_N$ = 535 K and the residual fraction of 12% staying paramagnetic down to the lowest temperatures. The lines are guides to the eye only. **b)** Zero field μSR spectra recorded at 2 K and 40 K. The long-lived oscillation proves a well defined internal magnetic field at the muon site indicating a long range ordered magnetic structure within the magnetic volume of the sample. The lines are fits to the data yielding internal magnetic fields (magnetic order parameters) at the muon site of 2.929(7) T and 2.937(7) T at 2 K and 40 K, respectively. **c)** Band structure of the antiferromagnetic band insulator Rb$_2$Fe$_4$Se$_5$. **d)** Schematic top-view of FeSe slabs with ordered iron vacancies. The 2D unit cell of Rb$_2$Fe$_4$Se$_5$ is shown.

Figure 3. **Superconductor. a,b)** Photoemission intensity in the superconducting state along the cuts running through the center and the corner of the BZ respectively. **c)** $k_F$-EDC with the peak at 8 meV binding energy corresponding to the superconducting gap. **d)** EDC dispersion of the electron-like states from panel **b** showing typical BCS bending-back behavior.

Figure 4. **Key features of the electronic structure. a-c)** Results of the band structure calculations for three iron-based superconductors along the cuts through the center **(a,c)** and corner of the BZ **(b)**. Bands are shown for each of the 10 $k_z$ values to emphasize the importance of $k_z$-dispersion. Equal weight is ascribed to each band. The places where the bands overlap are thus brighter according to the color scale used in Figs. 1,3. Fermi levels are shifted for a better agreement with the experiments. Insets schematically show the essential ingredients for high $T_c$ superconductivity. **d-f)** Corresponding Fermi surfaces. The $d_{xy}$ 2D sheet is omitted in **d** and **f**. The red from outside Fermi surfaces are electron-like, the green from outside – are hole-like. In panel **e** two Fermi surfaces are shown which correspond to the dashed lines in panel **b**. ARPES experiment shows that both types are observed in optimally-doped 122 materials [31].

**Figure 1**

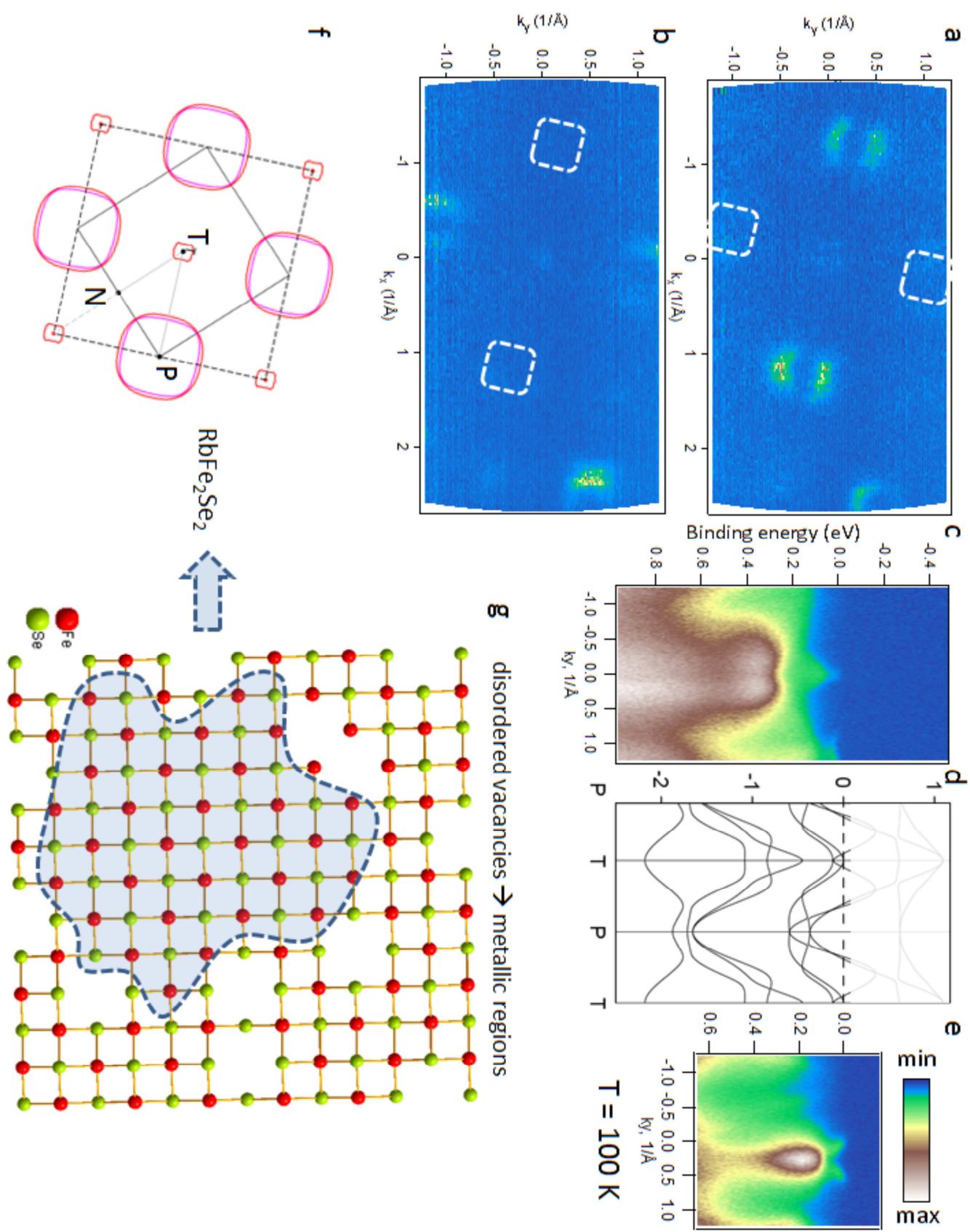

**Figure 2**

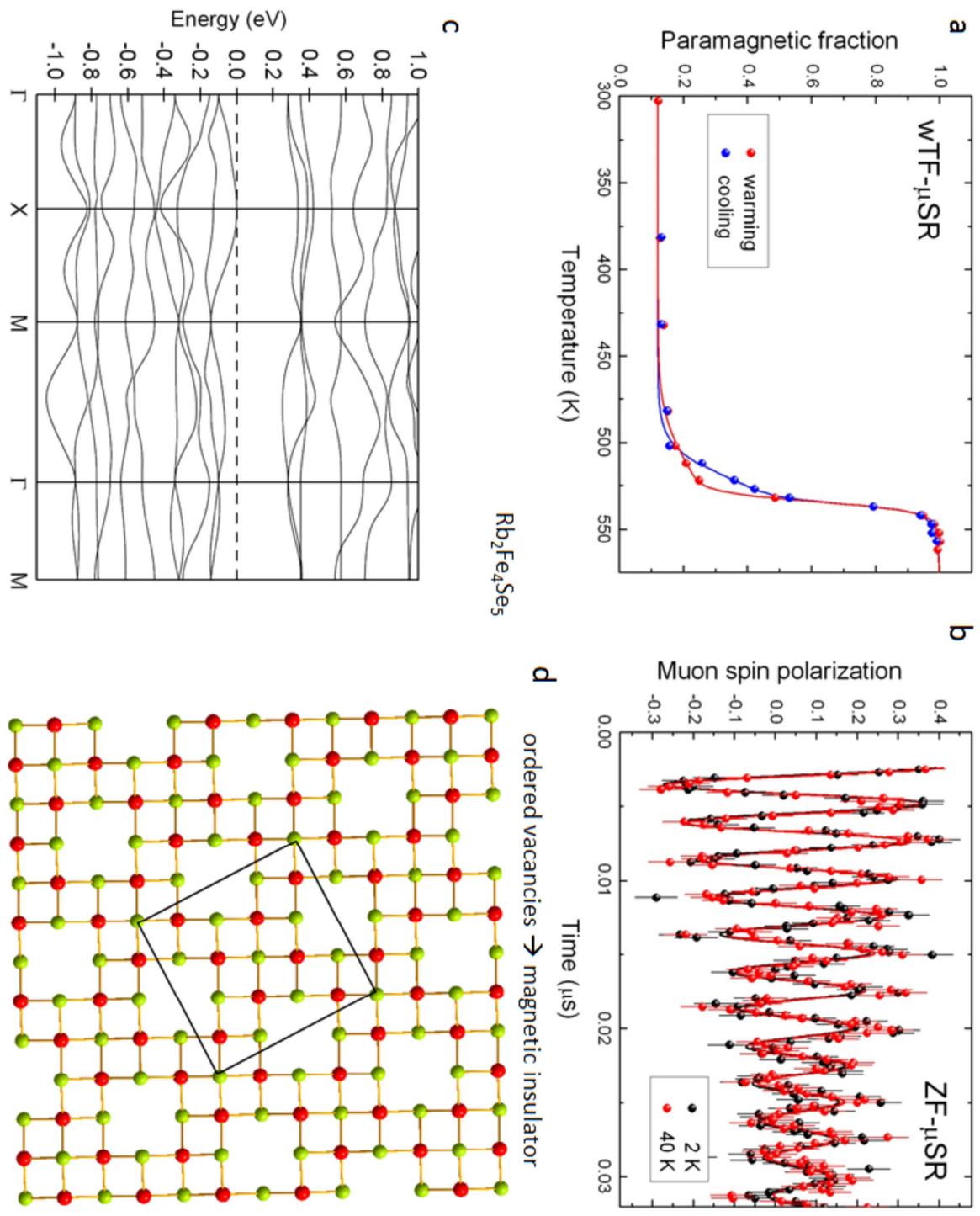

**Figure 3**

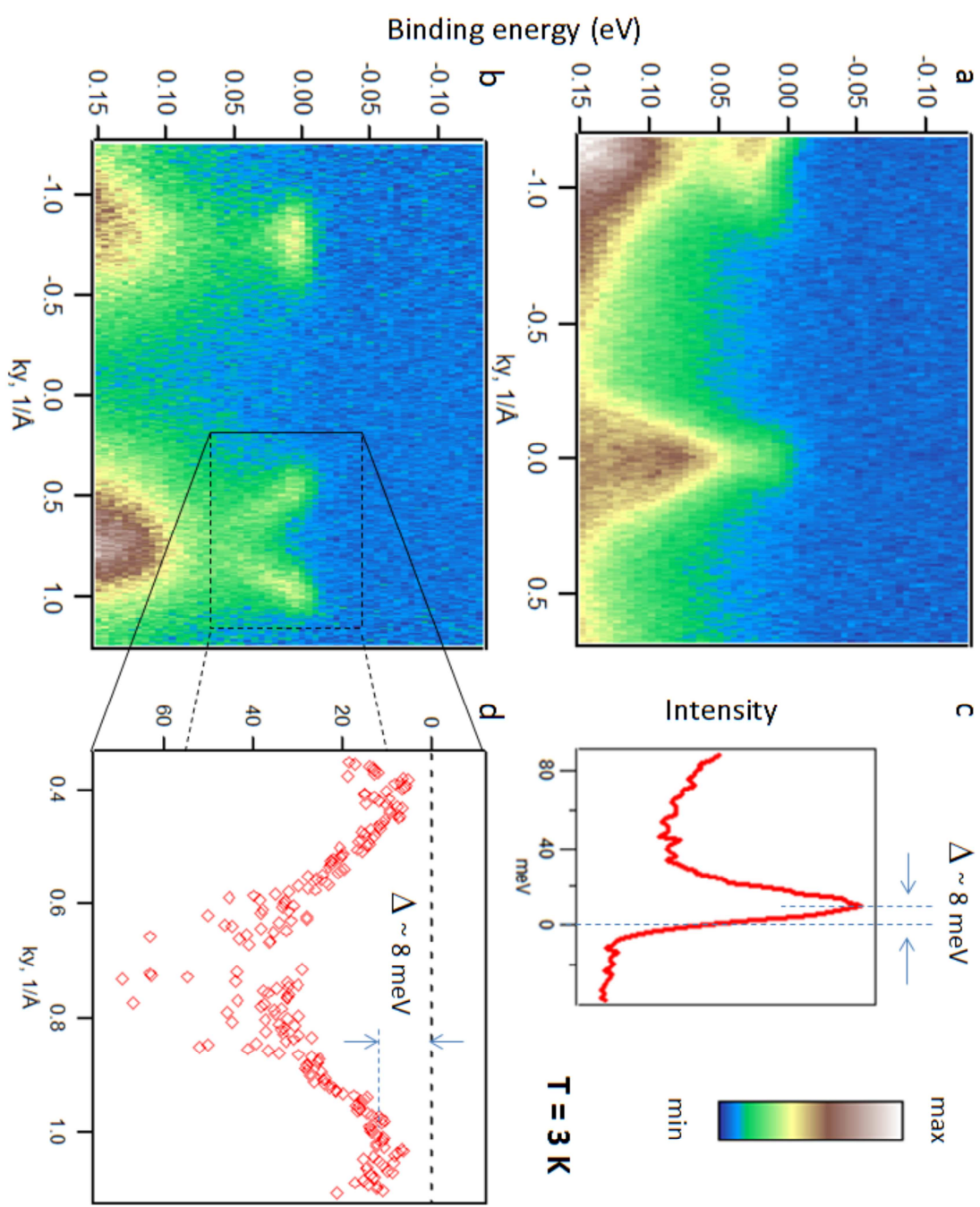

**Figure 4**

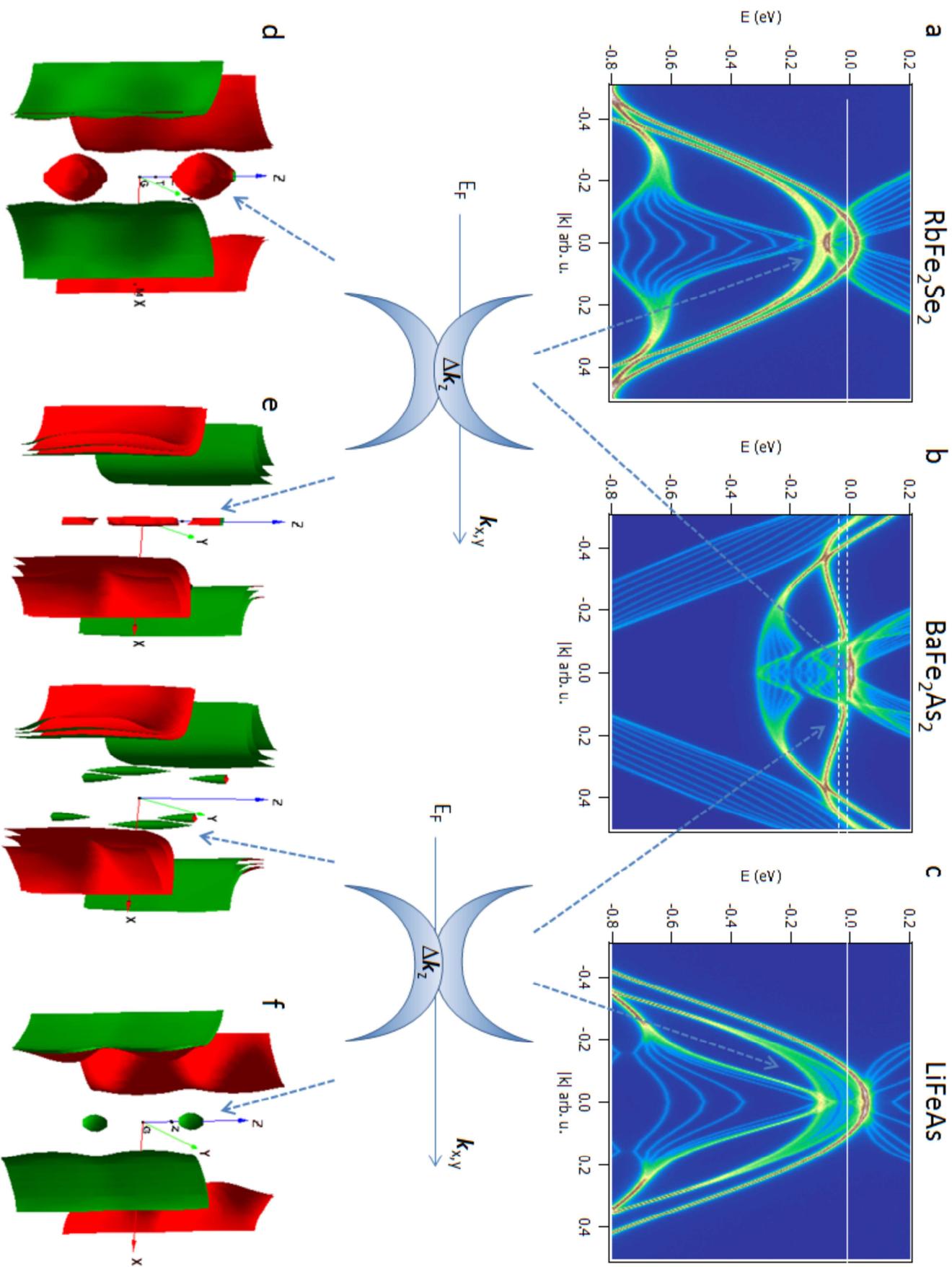